# Heat Capacity Anomaly of LSCO Cuprates with Nonmagnetic Impurity

M. M. Nadareishvili* K. A. Kvavadze

Andronikashvili Institute of Physics, Tbilisi, Georgia

*Corresponding author
E-mail address: m.nadareishvili@aiphysics.ge

## Abstract

The low temperature heat capacity investigations on pure(y=0) and Zn-doped $La_{1.84}Sr_{0.16}Cu_{1-y}Zn_yO_4$ samples have been performed in the interval of 1.8-60K by a new method of high-precision, pulsed differential calorimetry (PDC), which provides the measurements under the equilibrium conditions, in contrast to commonly used differential scanning calorimeters (DSC). For these systems a new heat capacity anomaly was observed, which is associated with Zn impurities and has the form of one wide peak. It does not show a phonon character shifts towards the higher temperatures with the increase of Zn content.

## 1.Introduction

The unsolved problem connected with the doping of HTSC cuprates by Zn is the magnetism of $CuO_2$ planes. In particular, though Zn is a nonmagnetic impurity, in the volume magnetization of HTSC cuprate $YBa_2Cu_3O_{7-\delta}$ at the doping with Zn, a Curie-like term with the efficient magnetic moment $P_{eff} \approx 1\mu_B$ is observed [1], which is connected with this impurity. NMR investigations of $^{63}Cu$ showed that, the mentioned effect is connected with the development of staggered magnetic moments induced by zinc and located in vicinity of each Zn site of Cu [2]. It was also shown that the contribution to the volume magnetization made by zinc can be explained by these induced magnetic moments. According to the earlier investigations of $YBa_2Cu_3O_{7-\delta}$ cuprates doped by zinc, the Curie law was fulfilled [1], giving the evidence of the absence of interaction among the efficient magnetic moments induced by zinc even at high concentrations of Zn, when the distance between the impurities was less than the typical diameter of staggered regions, that seemed surprising. More exact magnetic measurements using SQUIDs allowed to record the fulfillment of Curie-Weiss law with a low value of temperature θ [3], giving the evidence of the existence of interaction among these moments.

Thermodynamic investigations of HTSC cuprates were mainly carried out by differential scanning calorimeters (DSC) [4,5]. The differential method was used for reducing large phonon contributions to the experimentally measured difference in heat capacities of the sample and the reference, in order to make it possible to observe directly the fine thermal effects such as the uncompensated electron heat capacity and magnetic excitations. It should be noted that in DSC a continuous heating of samples is used and, hence, the measurements are made under the conditions being quasi-equilibrium. These investigations were mainly dealt with the study of the properties of HTSC electron heat capacity. The plots show well the jumps of electron heat capacity at superconducting transitions, as well as the existence of magnetic term in the heat capacity of Zn-doped yttrium cuprates $YBa_2(Cu_{1-x}Zn_x)_3O_7$ with the peak at about 5 K [4].

As for the investigations of HTSC heat capacity of cuprates by conventional heat pulse technique under the equilibrium conditions, they were carried out at the temperature lower than 10 K, as above 10 K the phonon contribution to the heat capacity of the sample becomes larger and makes it impossible to measure precisely the electron heat capacity and other "fine effects". In [6,7] at the introduction of Zn in $La_{2-x}Sr_xCuO_4$, the magnetic contribution was observed to the heat capacity of these samples, which, on the plots of the linear term of heat capacity γ (=$C_{el}$/T) was revealed as a bend upwards at the temperature lower than 3.5 K. The value of this anomaly increased with the increase of the concentration of Zn impurity.

**2. Experiment**

We have performed the measurement of low temperature heat capacity on the pure and Zn doped LSCO samples by the method of pulsed differential calorimetry (PDC). At present, the most sensitive and, hence, the most commonly used calorimeters are the differential scanning calorimeters (DSC). However, these devices have one significant disadvantage - they measure the difference in heat capacity between the studied sample and the reference sample under the non-equilibrium conditions, as the heat capacity difference was measured in the continuous heating regime. We have developed a new method of calorimetry, on the basis of which the high-sensitive differential calorimeter of a new type – Pulsed Differential Calorimeter was created [8-10]. This calorimeter can make the measurements with a very high sensitivity under equilibrium conditions, as the measurement of the heat capacity difference was carried out in the pulsed heating regime. It combines the high sensitivity of the modern, differential calorimeters of continuous heating type, and the high accuracy of the classical pulsed calorimeters. By this calorimeter the high precision measurements of the anomaly of low-temperature heat capacity were made in different substances [10,11].

The samples $La_{1.84}Sr_{0.16}Cu_{1-y}Zn_yO_4$ where fabricated at Sussex University by a standard solid-state reaction in which the initial combination of the constituent chemical compounds was made by solid state mixing, and then the samples were sintered at the temperature of 1030 K while being subjected to a high pressure. Subsequent measurements of the susceptibility showed that the samples with (x=0.16, y=0.033), (x=0.16, y=0.06), did not display the superconducting transitions for temperatures down to 4.2 K, being the lowest temperature in our measurements of susceptibility. On the other hand, the sample with (x=0.16, y=0.00) showed the superconducting transition at the temperature of 38 K.

### 3. Results and discussion

The heat capacity $C(y,T)$ of $La_{1.84}Sr_{0.16}Cu_{1-y}Zn_yO_4$ systems above 3.5 K is the sum of electron (hole) $C_{el}(y,T)$ and phonon $C_{ph}(y,T)$ contributions [6]. In its turn, $C_{el}(y,T) = \gamma(y)T + C_S(y,T)$, where $\gamma(y)$ is the coefficient of linear term of $C_{el}$, $C_S(y,T)$ is the heat capacity of superconducting electrons. As a result:

$$C(y,T) = \gamma(y)T + C_S(y,T) + C_{ph}(y,T) \qquad (1)$$

Difference in heat capacity between the sample under investigation and the reference sample with the different amount ($y_1, y_2$) of Zn is equal to

$$\Delta C(y_2,y_1,T) = \Delta\gamma(y_2,y_1)T + \Delta C_S(y_2,y_1,T) + \Delta C_{ph}(y_2,y_1,T) \qquad (2)$$

where

$$\Delta\gamma(y_2,y_1) = \gamma(y_2) - \gamma(y_1)$$
$$\Delta C_S(y_2,y_1,T) = C_S(y_2,T) - C_S(y_1,T)$$
$$\Delta C_{ph}(y_2,y_1,T) = C_{ph}(y_2,T) - C_{ph}(y_1,T)$$

Substitution of Zn for Cu does not affect the phonon heat capacity, as the masses and ionic radiuses of Zn and Cu are equal, and thus we can consider that $C_{ph}(y_2,T) - C_{ph}(y_1,T) = 0$ [6]. If the sample under investigation is superconducting with $y_2=0$ and the reference sample is non-superconducting with $y_1 = y_c$ ($y_c=0.033$ is the critical (minimum) concentration of Zn, when superconductivity is suppressed), i.e. $C_S(y_1,T)=0$, and Exp. 2 gives

$$\Delta C(0,y_1,T) - \Delta\gamma(0,y_1)T = C_S(0,T) \qquad (3)$$

Using the high-precision PDC technique, the heat capacity difference $\Delta C(0,y_1,T)$ between the superconducting sample under investigation $La_{1.84}Sr_{0.16}CuO_4$ ($y=0$) and the non-superconducting reference sample $La_{1.84}Sr_{0.16}Cu_{0.967}Zn_{0.033}O_4$ ($y=0.033$) was measured in the low temperature interval of 1.8 – 60 K under the equilibrium conditions. The difference between the coefficients of linear term $\Delta\gamma(0,y_1)$ was estimated in a usual way by plotting the generally used relation $\Delta C(0,y_1,T)/T = f(T^2)$ [6] on the basis of the experimental data $\Delta C(0,y_1,T)$ in the 3.5 – 8 K temperature interval. It was found that $|\Delta\gamma(0,y_1)| = 8.4$ mj/mol. $K^2$.

Curve 2 in Fig. 1 (open circles) shows $\Delta C(0,y_1,T) - \Delta\gamma(0,y_1)T$ dependence for molar heat capacity difference between $La_{1.84}Sr_{0.16}CuO_4$ and $La_{1.84}Sr_{0.16}Cu_{0.967}Zn_{0.033}O_4$ ($y_1=0.033$ reference sample). One can note the appearance of the unphysical (negative) region for $C_S(0,T)$. To clarify the situation, we measured the heat capacity difference $\Delta C(y_2,y_1,T)$ between two non-superconducting samples with different content of Zn ($y_2=0.033$ - the investigated sample and $y_1=0.06$ - the reference sample). As the both samples are non-superconducting, $C_S(y,T) = 0$ and, if Exp. 1 presenting the heat capacity is valid, Exp. 2 shows that $\Delta C(y_2,y_1,T) - \Delta\gamma(y_2,y_1)T$ difference should be zero. However, as Fig. 2 shows, this dependence is of a complex form. Thus, it is evident that the representation of the heat capacity in the form of sum (1) is not complete and needs some correction, namely, the introduction of excess $\delta C(y,T)$ contribution. Hence, in general case, the heat capacity of Zn-doped LSCO samples above 3.5 K can be presented as follows

$$C(y,T) = \gamma(y)T + C_S(y,T) + C_{ph}(y,T) + \delta C(y,T) \qquad (4)$$

with $\delta C(0,T)=0$.

Then, the difference in heat capacity between the investigated sample $La_{1.84}Sr_{0.16}Cu_{0.967}Zn_{0.033}O_4$ and the reference sample $La_{1.84}Sr_{0.16}Cu_{0.94}Zn_{0.06}O_4$ will have the following form

$$\Delta C(y_2,y_1,T) - \Delta\gamma(y_2,y_1)T = \delta C(y_2,T) - \delta C(y_1,T) \qquad (5)$$
$$y_2 = 0.033 \text{ and } y_1 = 0.06$$

As Fig. 2. shows, in the normal phase (when $y \geq y_c$), $\delta C(y,T)$ has the form of a wide peak increasing almost linearly and shifting to high temperatures with the increase of zinc concentration. As $\delta C(y_1,T)$ is shifted according to the temperature relative to $\delta C(y_2,T)$, on the plot in Fig. 2. we have a valley according to (5). The experimentally observed shift of the maximum of $\delta C(y,T)$ value according to the temperature with the increase of impurity concentration is not characteristic of the anomalies of phonon part of heat capacity of materials [12], just on the contrary, it is in agreement with the magnetic nature of this contribution, which is characterized by a strong shift of the maximum according to the temperature at the increase of impurity concentration.

Taking into account the additional term in heat capacity, for $\Delta C - \Delta\gamma T$ value (shown in Fig.1, curve 2) from (4) we obtain the following expression

$$\Delta C(0,y_1,T) - \Delta\gamma(0,y_1)T = C_S(0,T) - \delta C(y_1,T), \qquad (6)$$

which explains the reason of appearance of the negative region on curve 2, Fig.1, and Exp. 3 will return to

$$\Delta C(0,y_1,T) - \Delta\gamma(0,y_1)T + \delta C(y_1,T) = C_S(0,T) \qquad (7)$$

Fig. 1, curve 1 (full circles) shows the dependence of $\Delta C(0,y_1,T) - \Delta\gamma(0,y_1)T + \delta C(y_1,T)$ on T with the excess $\delta C(y_1,T) = \delta C(0.033,T)$ contribution. $\delta C(0.033,T)$ was taken from Fig.2 (left peak). One can easily see that, there is not an unphysical region for $C_S(0,T)$.

## 4. Conclusion

Summing up the above-said, one can conclude that in LSCO ceramic superconductors at the introduction of Zn in non-superconducting state a new anomaly of low-temperature heat capacity arises, the value of which increases almost proportionally to Zn concentration and is shifted to high temperatures.

The modification of PDC calorimeter was supported by the INTAS Grant N1010-CT93-0046.

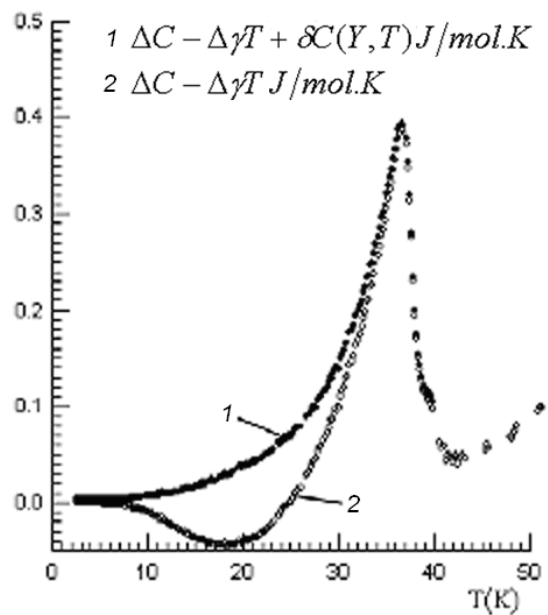

Fig. 1.

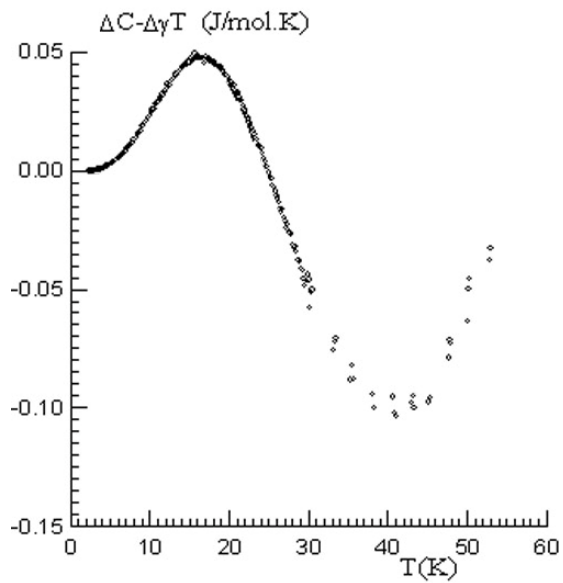

Fig. 2.

**Figure captions**

Fig. 1. Temperature dependence of the $\Delta C - \Delta\gamma T$ for molar heat capacity between $La_{1.84}Sr_{0.16}CuO_4$ and $La_{1.84}Sr_{0.16}Cu_{0.967}Zn_{0.033}O_4$ samples.
Curve 1 – with additional term $\delta C$, Curve 2 – without additional term $\delta C$.

Fig. 2. Temperature dependence of the $\Delta C - \Delta\gamma T$ for molar heat capacity between non-superconducting samples $La_{1.84}Sr_{0.16}Cu_{0.967}Zn_{0.033}O_4$ and $La_{1.84}Sr_{0.16}Cu_{0.94}Zn_{0.06}O_4$.